\documentclass[twocolumn,twoside,slac_two]{revtex4}
\usepackage{graphicx}
\usepackage{fancyhdr}
\newcommand{\beq}{\begin{equation}}
\newcommand{\eeq}{\end{equation}}
\newcommand{\ba}{\begin{array}}
\newcommand{\ea}{\end{array}}
\newcommand{\beqa}{\begin{eqnarray}}
\newcommand{\eeqa}{\end{eqnarray}}

\newcommand{\gsim}{\stackrel{>}{_\sim}}

\newcommand{\cO}{{\cal O}}
\newcommand{\cB}{{\cal B}}
\newcommand{\cL}{{\cal L}}

\newcommand{\no}{\nonumber}

\newcommand{\yuk}{{Y}}

\newcommand{\kppn}{K^+\rightarrow\pi^+\nu\bar\nu}
\newcommand{\klpn}{K_L\rightarrow\pi^0\nu\bar\nu}
\newcommand{\kpnn}{K \rightarrow\pi\nu\bar\nu}

\newcommand{\kLpee}{K_L \rightarrow\pi^0 e^+e^-}
\newcommand{\kLpmm}{K_L \rightarrow\pi^0 \mu^+\mu^-}

\newcommand{\sttop}{{\tilde t}}

\newcommand{\charg}{{\tilde \chi}}
\newcommand{\LLN}{{\Lambda_{\rm LN}}}
\newcommand{\LLFV}{{\Lambda}}

\def\npb#1#2#3{    {\it Nucl. Phys.}~B {\bf #1}, #3 (#2)}
\def\plb#1#2#3{    {\it Phys. Lett.}~B {\bf #1}, #3 (#2)}

\def\prd#1#2#3{    {\it Phys. Rev.}~D {\bf #1}, #3 (#2)}

\def\epjc#1#2#3{   {\it Eur. Phys. J.}~C {\bf #1}, #3 (#2)}
\def\jhep#1#2#3{   {\it JHEP  }{\bf #1}, #3 (#2)}

\begin{document}

\title{Flavor Physics with light quarks and leptons}

%

\author{Gino Isidori}
\affiliation{
INFN, Laboratori Nazionali di Frascati, Via E. Fermi 40, I-00044 Frascati, Italy }

\begin{abstract}
The impact of rare $K$ decays and lepton-flavor 
violating processes in shedding light on physics beyond the Standard Model
is reviewd. To this purpose, I first recall the formulation of 
the Minimal Flavor Violation hypothesis
--both in the quark and in the lepton sector--
and then use it as guiding principle 
in comparing the new-physics 
sensitivity of different rare processes. 
On the phenomenological side, the discussion is focused mainly 
on the impact of $K\to \pi\nu\bar\nu$, 
$\mu\to e \gamma$, and lepton-flavor universality tests
with $K_{\ell 2}$ and $\pi_{\ell 2}$ decays.
\end{abstract}

\maketitle

\thispagestyle{fancy}


\section{Introduction}
Despite the great phenomenological success of the Standard Model (SM),
it is natural to consider this theory only as the low-energy limit
of a more general model. More precisely, the SM Lagrangian can be 
regarded as the renormalizable part of an effective field theory (EFT), 
valid up to some still undetermined cut-off scale $\Lambda$ ($\Lambda > M_W$). 
Since the SM is renormalizable, we have no clear clues 
about the value of $\Lambda$; however, 
theoretical arguments based on a natural solution of the 
hierarchy problem suggest that $\Lambda$ should not exceed a few TeV. 
A key experimental test of this hypothesis will soon be 
performed at the LHC, with the direct exploration 
of the TeV energy scale.

The direct search for new degrees of freedom at the TeV 
scale (the so called {\em high-energy frontier}) is not 
the only tool at our disposal to shed light on physics 
beyond the SM. A complementary and equally important 
source of information about the underlying theory
is provided by high-precision low-energy experiments 
(the so called {\em high-intensity frontier}).
The latter are particularly interesting in determining 
the symmetry properties of the new degrees of freedom. 
As I will discuss in this talk, flavor-changing neutral-current (FCNC) 
transitions in $K$ and $\mu$ decays, 
and lepton-flavor (LF) universality tests in $K_{\ell 2}$ and $\pi_{\ell 2}$
decays, offer a unique opportunity to study the 
flavour structure of physics beyond the SM.

\subsection{The flavor problem}

As long as we are interested only in low-energy experiments, 
the EFT approach to physics beyond the SM is particularly 
useful. It allows us to analyse all realistic extensions 
of the model in terms of few unknown parameters (the 
coefficients of the higher-dimensional operators 
suppressed by inverse powers of $\Lambda$) and 
to compare the  new-physics (NP) sensitivity of different
low-energy observables. 

The non-renormalizable 
operators should naturally induce large effects 
in processes which are not mediated by tree-level SM amplitudes, 
such as FCNC processes. 
Up to now there is no evidence of these effects and this 
implies severe bounds on the effective scale of dimension-six 
FCNC operators. For instance, the good agreement between SM 
expectations and experimental determinations of $K^0$--${\bar K}^0$ 
mixing leads to bounds above $10^4$~TeV for the effective scale 
of $\Delta S=2$ operators, i.e.~well above the few TeV 
range suggested by the Higgs sector. Similar bounds are 
obtained for the scale of LF-violating operators 
contributing to FCNC transitions in the lepton sector, 
such as $\mu\to e\gamma$.

The apparent contradiction between these 
two determinations of  $\Lambda$ is a manifestation of what in 
many specific frameworks (supersymmetry, techincolour, etc.)
goes under the name of {\em flavor problem}:
if we insist with the theoretical prejudice that new physics has to 
emerge in the TeV region, we have to conclude that the new theory 
possesses a highly non-generic flavor structure. 
Interestingly enough, this structure has not been clearly identified yet,
mainly because the SM, i.e.~the low-energy 
limit of the new theory, doesn't possess an exact flavor symmetry.

The most reasonable (but also most {\em pessimistic}) solution
to the flavor problem is the so-called 
{\it Minimal Flavor Violation} (MFV) hypothesis \cite{Georgi,MFV,MLFV,MFV2}. 
Under this assumption, which will be discussed  
in detail in the next sections, flavor-violating 
interactions are linked to the
known structure of Yukawa couplings also beyond the SM. 
As a result, non-standard contributions in FCNC 
transitions turn out to suppressed to a level consistent 
with experiments even for $\Lambda \sim$~few TeV.

On the most interesting aspects of the MFV hypothesis 
is the possibility to formulate it within the  
general EFT approach to physics beyond the SM \cite{MFV,MLFV}. 
The effective theories based on this symmetry principle
allow us to establish unambiguous correlations 
among NP effects in different FCNC transitions. 
These falsifiable predictions are the key ingredient   
to identify in a model-independent way which are the 
irreducible sources of breaking of the flavor symmetry.

\section{MFV in the quark sector}
The pure gauge sector of the SM is invariant under
a large symmetry group of flavor transformations: 
$G_F \equiv {\rm SU}(3)^3_q \otimes  {\rm SU}(3)^2_\ell\otimes U(1)^5$,
where 
${\rm SU}(3)^3_q  
= {\rm SU}(3)_{Q_L}\otimes {\rm SU}(3)_{U_R} \otimes {\rm SU}(3)_{D_R}$,
    ${\rm SU}(3)^2_\ell 
=  {\rm SU}(3)_{L_L} \otimes {\rm SU}(3)_{E_R}$
and three of the five $U(1)$ charges can be identified with 
baryon number, lepton number and hypercharge \cite{Georgi,MFV}. 
This large group and, particularly the ${\rm SU}(3)$ 
subgroups controlling flavor-changing transitions, is 
explicitly broken by the Yukawa interaction
\beq
\cL_Y  =   {\bar Q}_L \yuk_D D_R  H
+ {\bar Q}_L {\yuk_U} U_R  H_c
+ {\bar L}_L {\yuk_E} E_R  H {\rm ~+~h.c.}
\label{eq:LY}
\eeq
Since $G_F$ is broken already within the SM, 
it would not be consistent to impose it as an exact symmetry 
of the additional degrees of freedom
present in SM extensions: even if absent a the tree-level,
the breaking of $G_F$ would reappear at the quantum level 
because of the Yukawa interaction.  
The most restrictive hypothesis 
we can make to {\em protect} the breaking of $G_F$ 
in a consistent way, is to assume that 
$\yuk_D$, $\yuk_U$ and $\yuk_E$ are the only source of 
$G_F$-breaking also beyond the SM.

To implement and interpret this hypothesis in a natural way, 
we can assume that $G_F$ is indeed a good symmetry, promoting 
the $\yuk$ to be dynamical fields with 
non-trivial transformation properties under $G_F$:
\beqa
&&\yuk_U \sim (3, \bar 3,1)_{{\rm SU}(3)^3_q}~,\qquad
\yuk_D \sim (3, 1, \bar 3)_{{\rm SU}(3)^3_q}~,\no \\
&&\yuk_E \sim (3, \bar 3)_{{\rm SU}(3)^2_\ell}~.
\eeqa
If the breaking of $G_F$ occurs at very high energy scales 
 --well above the TeV region where the we expect new degrees--
at low-energies we would only be sensitive to the background values of 
the $\yuk$, i.e. to the ordinary SM Yukawa couplings. 
Employing the EFT language, 
we then define that an effective theory satisfies the criterion of
Minimal Flavor Violation if all higher-dimensional operators,
constructed from SM and $\yuk$ fields, are (formally)
invariant under the flavor group $G_F$ \cite{MFV}. 

According to this criterion, one should in principle 
consider operators with arbitrary powers of the (adimensional) 
Yukawa fields. However, a strong simplification arises by the 
observation that all the eigenvalues of the Yukawa matrices 
are small, but for the top one, and that the off-diagonal 
elements of the CKM matrix ($V_{ij}$) are very suppressed. 
It is then easy to realize that, similarly to the pure SM case, 
the leading coupling ruling all FCNC transitions 
with external down-type quarks is:
\beq
\label{eq:FC}
(\lambda_{\rm FC})_{ij} = \left\{ \ba{ll} \left( \yuk_U \yuk_U^\dagger \right)_{ij}
\approx \lambda_t^2  V^*_{3i} V_{3j}~ &\qquad i \not= j~, \\
0 &\qquad i = j~. \ea \right.
\eeq
As a result, within this framework the bounds on 
the scale of dimension-six FCNC effective operators 
turn out to be in the few TEV range (see Ref.~\cite{UTfit}
for updated values). Moreover, the flavour structure 
of (\ref{eq:FC}) implies a well-defined link among 
possible deviations from the SM in FCNC transitions 
of the type $s\to d$, $b\to d$, and  
$b\to s$.\footnote{~Within the MFV framework, 
these three types of FCNC processes 
are the only quark-level transitions where 
observable deviations from 
the SM are expected.}

The idea that the CKM matrix rules the strength of FCNC 
transitions also beyond the SM has become a very popular 
concept in the recent literature and has been implemented 
and discussed in several works (see e.g.~Ref.~\cite{MFV2}). 
However, it is worth stressing that the CKM matrix 
represent only one part of the problem: a key role in
determining the structure of FCNCs is also played  by quark masses
(via the GIM mechanism), or by the Yukawa eigenvalues. 
In this respect the above MFV criterion provides the maximal protection 
of FCNCs (or the minimal violation of flavor symmetry), since the full 
structure of Yukawa matrices is preserved. We finally emphasize that, 
contrary to other approaches, the above MFV criterion 
is based on a renormalization-group-invariant symmetry argument,
which is independent from the specific new-physics framework.

\section{Rare $K$ decays}
In the kaon sector is often difficult to control long-distance 
effects to a level sufficient to perform precise test of 
short-distance dynamics. This happens in the four 
golden modes $\klpn$, $\kppn$, $\kLpee$, and $\kLpmm$,
which represent a unique window on $s\to d$ FCNC transitions. 
As shown in Table~\ref{tab:rareK},
the theoretical cleanness of these 
four modes is not the same. The two neutrino channels are 
exceptionally clean: their decay rates can be computed to 
a degree of precision not matched by any other FCNC process 
in the $B$ and $K$ systems \cite{Haisch}.

\begin{table}[t]
\begin{center}
\begin{tabular}{|c|c|c|c|}
\hline         & short-distance  &  irreducible     &  SM BR    \\
       Channel & contribution    &  th. error on    & (central  \\
               & (rate \%)       &  s.d. contrib.   &   value)  \\  \hline 
$\klpn$   &  $> 99\% $ & $\sim 1\%$   & $3 \times 10^{-11}$ \\  \hline 
$\kppn$   &  $88\% $ & $\sim 3\%$   & $8 \times 10^{-11}$ \\  \hline
$\kLpee$  &  $38\% $ & $\sim 15\%$  & $3.5 \times 10^{-11}$ \\  \hline
$\kLpmm$  &  $28\% $ & $\sim 30\%$  & $1.5 \times 10^{-11}$ \\  \hline
\end{tabular}
\caption{\label{tab:rareK} Summary of the short-distance sensitivity of the 
four most-interesting rare $K$ decays. The second column denotes the 
contribution to the total rate determined by electroweak dynamics (top-quark loops). 
The third column indicates the irreducible (non-parametric) error on the 
short-distance amplitude, as extracted from the corresponding rate
measurement (in the limit of negligible exp. error).}
\end{center}
\end{table}

\subsection{SM predictions}
$\underline{K\to \pi \nu\bar\nu}\ $
The main reason for the exceptional theoretical cleanness of 
$\kpnn$ decays is the fact that --within the SM-- 
these processes are mediated by electroweak amplitudes 
of $O(G_F^2)$ which exhibit a power-like GIM 
mechanism and are largely dominated by top-quark loops. 
This property implies a severe 
suppression of non-perturbative effects~\cite{CPC,Falk,IMS}.
By comparison, it should be noted that typical 
loop-induced amplitudes relevant to meson decays 
are of $O(G_F\alpha_s)$ (gluon penguins) or 
$O(G_F\alpha_{\rm em})$ (photon penguins), 
and have only a logarithmic-type GIM mechanism
(which implies a much less severe suppression 
of long-distance effects).

A related important virtue,
is the fact that the leading contributions to 
$K\to\pi\nu\bar\nu$ amplitudes can be described 
in terms of a single dimension-six effective operator,
\beq
Q_{sd}^{\nu\nu} = \bar{s} \gamma^\mu  (1-\gamma_5) d ~ \bar{\nu} \gamma_\mu   (1-\gamma_5) \nu~,
\label{eq:QnL}
\eeq
both in the SM and in MFV models. 
The hadronic matrix elements of  $Q_{sd}^{\nu\nu}$
relevant to $K\to\pi\nu\bar\nu$ amplitudes 
can be extracted directly from the well-measured 
$K \to\pi e\nu$ decays, including isospin 
breaking corrections~\cite{MP}. 

\begin{figure}[t]
\begin{center}
\includegraphics[width=82mm]{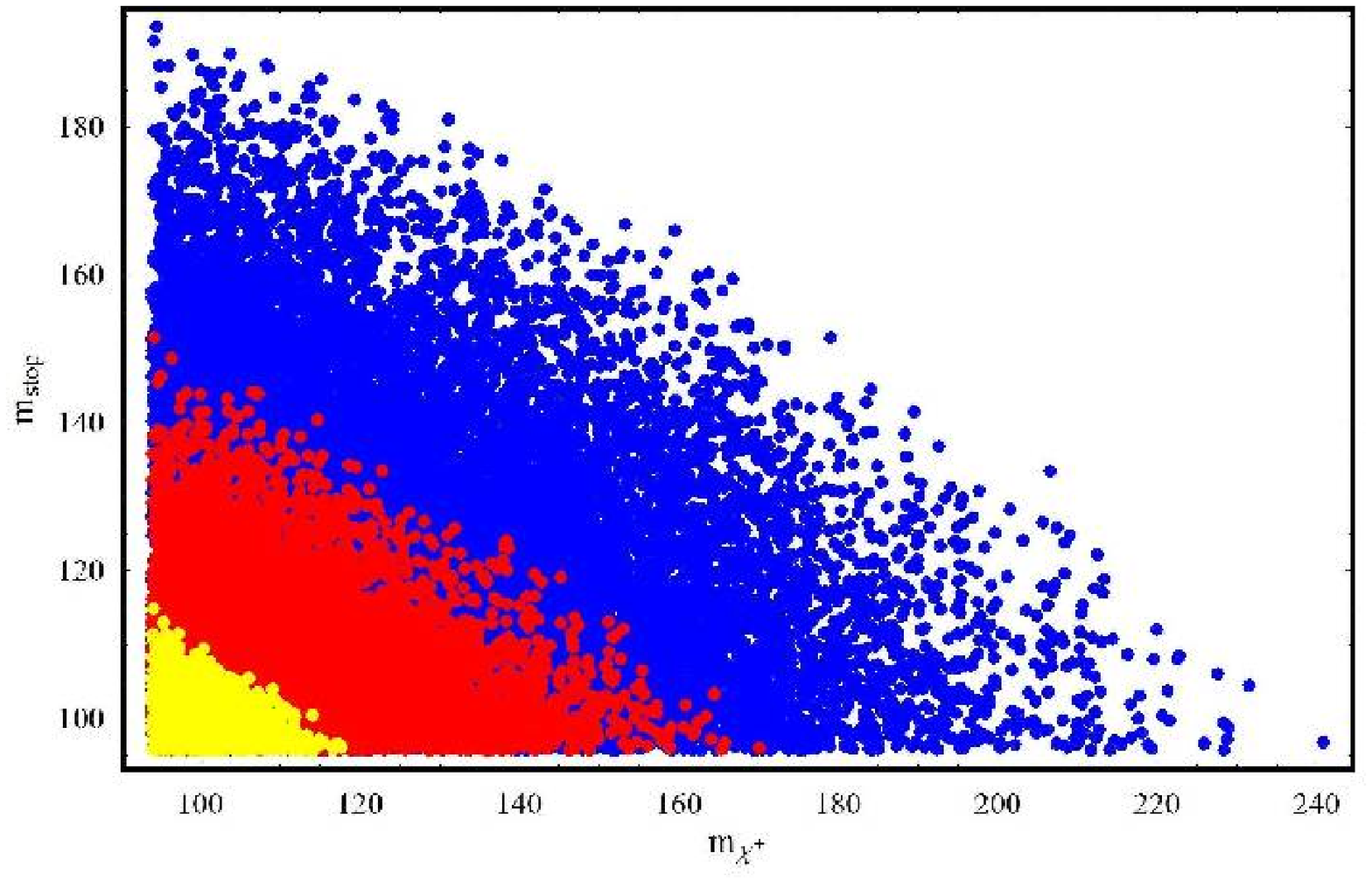}
\caption{\label{fig:stopvschar} Regions in the
$m_\sttop$ -- $m_\charg$ plane (lightest stop and chargino masses)
allowing enhancements of $\cB(\kppn)$ 
of more than $11\%$ (yellow/light gray),  $8.5\%$ (red/medium gray) 
and $6\%$ (blue/dark gray) in the MFV-MSSM scenario \cite{IMPST}, for $\tan\beta=2$
and $M_{H^+} > 1$~TeV [the corresponding 
enhancements for $\cB(\klpn)$  are 15\%, 12.5\% and 10\%, 
respectively].}
\end{center}
\end{figure}

The dominant theoretical error in estimating 
the $K^+\to\pi^+ \nu\bar{\nu}$ rate within the SM 
is due to the subleading, but non-negligible, charm contribution. 
A significant step forward in the reduction of this error has 
recently been achieved in Ref.~\cite{NNLO}, where the 
charm contribution to the Wilson coefficient of $Q_{sd}$
has been evaluated at the NNLO accuracy in QCD.
Thanks to this work, the intrinsic theoretical error 
in the perturbative charm contribution turns out 
to be safely negligible: the dominant 
uncertainty  is the parametric error induced by the 
knowledge of $m_c$~\cite{NNLO}, which induces  
a $\sim \pm 5\%$ error in the total rate. 
Recently,
the error associated to non-perturbative effects 
around and below the charm scale (dimension-eight 
operators and light-quark loops) has also been 
quantified and reduced~\cite{IMS}: this
residual uncertainty induces a $\sim \pm 3\%$ error
on the rate. As shown in Ref.~\cite{IMT},
this error could possibly be reduced in the 
future by means of appropriate lattice calculations.
Putting all the ingredients together, taking into 
account the sizable parametric uncertainty on the 
CKM matrix elements, the present updated prediction 
for the charged channel reads:
\beq
\cB(K^+\to\pi^+\nu\bar\nu) = (8.2 \pm 1.0) \times 10^{-11} 
\eeq

The case of $K_L\to\pi^0 \nu\bar{\nu}$ is even 
cleaner from the theoretical point of view \cite{Litt}.
The CP structure of $Q_{sd}$ implies that 
only the CP-violating part of the dimension-six effective 
Hamiltonian  --where the charm contribution is absolutely negligible--
contributes to  $K_2 \to\pi^0 \nu\bar{\nu}$. As a result, 
the dominant direct-CP-violating component 
of the $K_L \to\pi^0 \nu\bar{\nu}$ amplitude is completely 
saturated by the top contribution (which receives tiny  QCD corrections).
Intermediate and long-distance effects in this process
are confined only to the indirect-CP-violating 
contribution \cite{BB3} and to the CP-conserving one \cite{CPC},
which are both extremely small. This allows us to write an
expression for the $K_L\to\pi^0 \nu\bar{\nu}$ rate in terms of 
short-distance parameters, namely
\beqa
&& {\cal B}(K_L\to\pi^0 \nu\bar{\nu})_{\rm SM} ~=~ 4.16 \times 10^{-10} \no \\
&& \ \times \left[
\frac{\overline{m}_t(m_t) }{ 167~{\rm GeV}} \right]^{2.30} \left[ 
\frac{\Im(V^*_{ts} V_{td})}{ \lambda^5 } \right]^2~, 
\eeqa
which has a theoretical error below $3\%$.

\begin{figure}[t]
\begin{center}
\vspace{0.2 cm}
\includegraphics[scale=0.28,angle=-90]{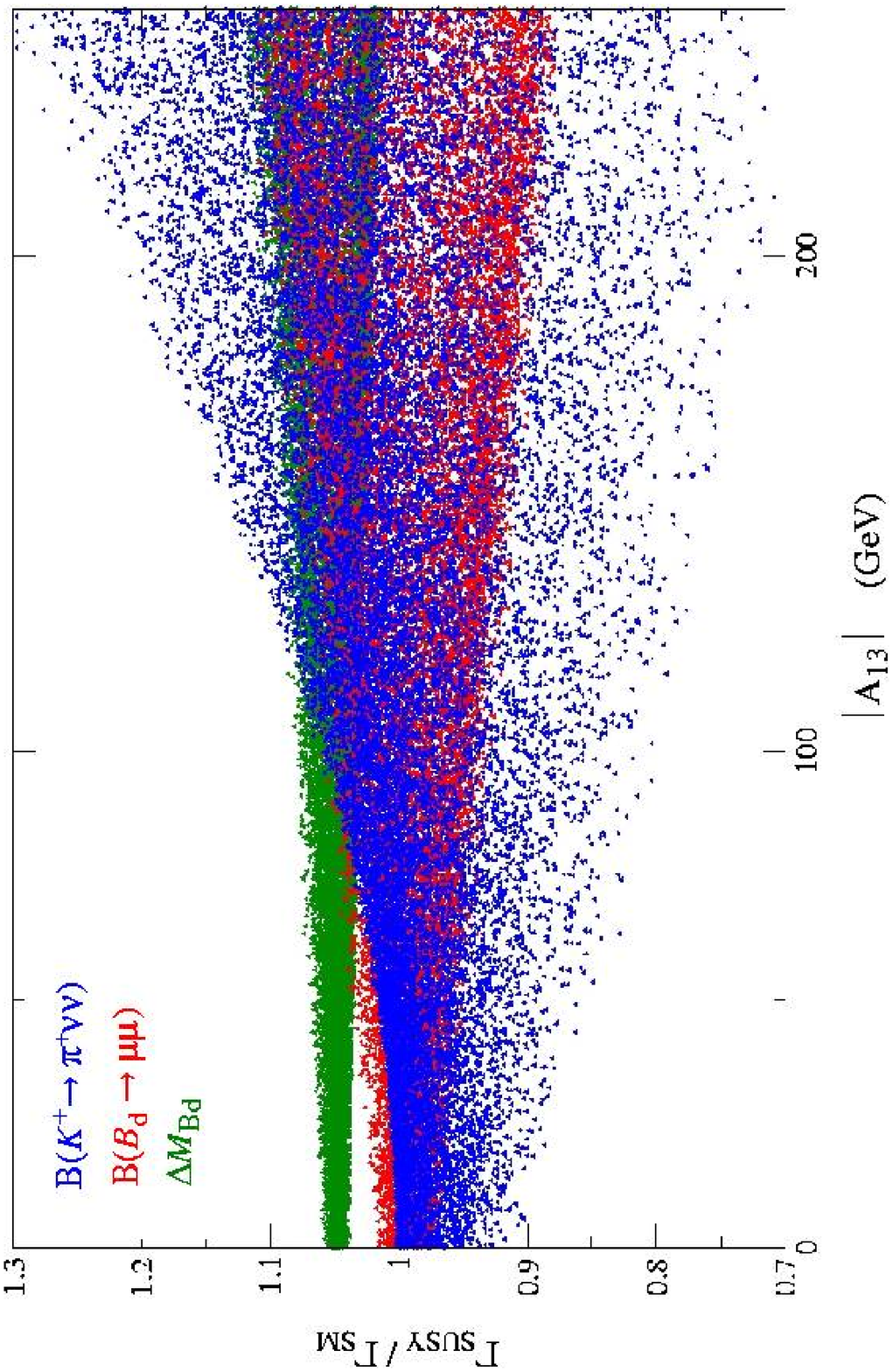} \\
\vspace{1.0 cm}
\includegraphics[scale=0.28,angle=-90]{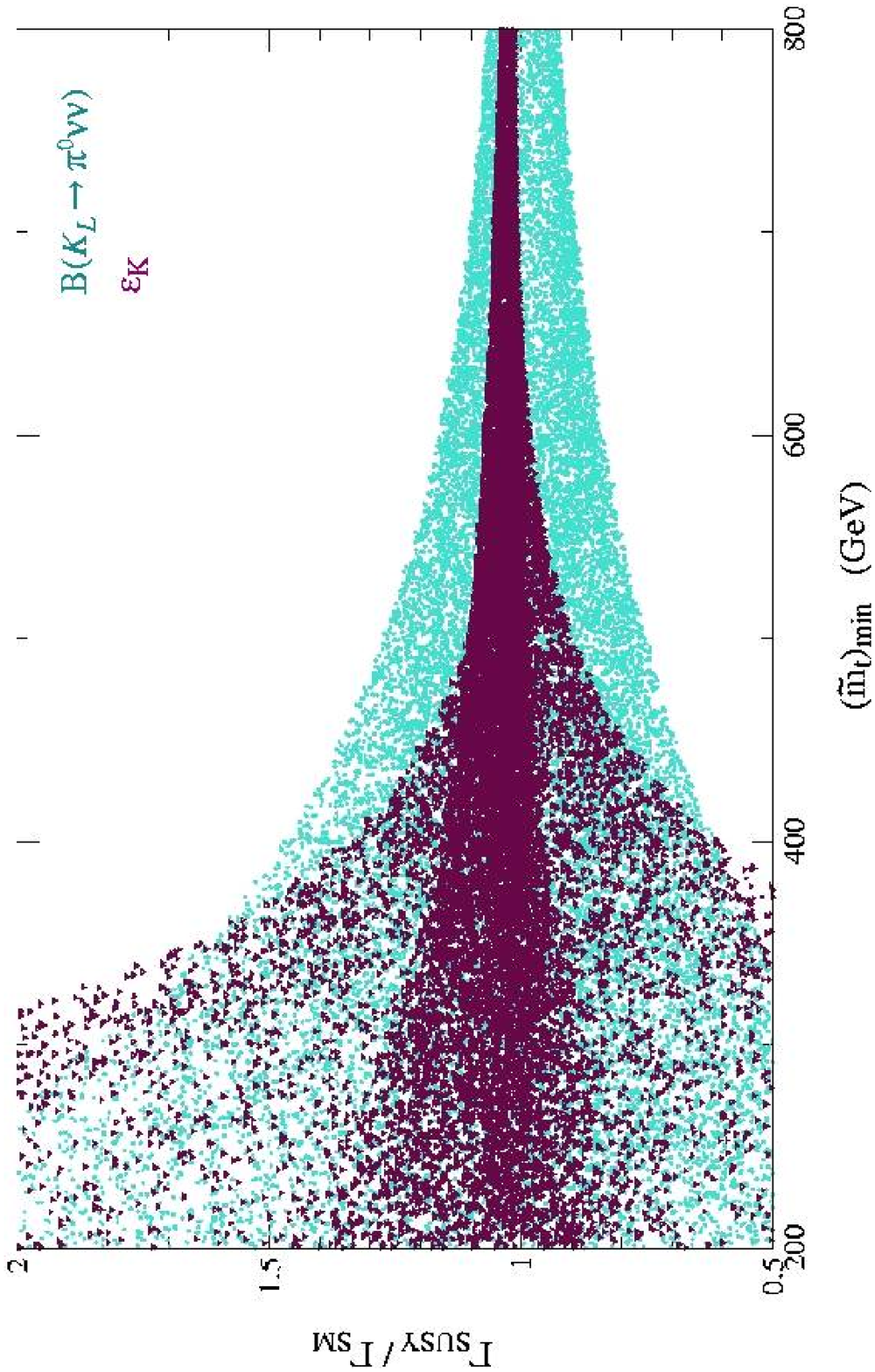} \\
\vspace{0.5 cm}
\caption{\label{fig:Kpnn} Dependence of various FCNC 
observables (normalized to their SM value) on the up-type trilinear terms ($A_{13}$) 
and on squark masses in the general MSSM \cite{IMPST}
Upper plot: $\cB(\kppn)$ (blue/dark gray), $\cB(B_d\to\mu^+\mu^-)$ (red/gray lower-region), 
$\Delta M_{B_d}$ (green/gray upper-region) as a function of $A_{13}$.
Lower plot: $\epsilon_K$ (bordeaux/dark gray) and $\cB(\klpn)$ (light blue/light gray)
as a function of  the lightest up-type squark mass. }
\end{center}
\end{figure}

$\underline{K\to \pi \ell^+\ell^-}\ $
The  GIM mechanism of the $s \to d \gamma^*$ amplitude 
is only logarithmic.  
As a result, the  $K \to \pi \gamma^* \to \pi \ell^+\ell^-$ amplitude 
is completely dominated by long-distance dynamics 
and  provides a large contribution to the CP-allowed transitions  
$K^+ \to \pi^+ \ell^+ \ell^-$ and $K_S \to \pi^0 \ell^+ \ell^-$.
  
The situation is very different for the very-suppressed 
$K_L \to \pi^0 \ell^+ \ell^-$ modes. 
The decay amplitudes of these
processes have three main ingredients: i)~a clean direct-CP-violating
component determined by short-distance dynamics; ii)~an indirect-CP-violating 
term due to $K_{L}$-$K_{S}$ mixing and the CP-allowed $K_S \to \pi^0\gamma^*$
transition; iii)~a
long-distance CP-conserving component due to two-photon intermediate
states. Although generated by very different dynamics, these three components
are of comparable size within the SM. The precise knowledge
about their magnitude and sign (particularly of ii and iii) 
has substantially improved in the last few years \cite{BDI,ISU,EdR}. 
This improvement has been made possible by the 
observation of the $K_{S} \rightarrow\pi^{0}
\ell^{+} \ell^{-}$ decays \cite{NA48ll} and also by precise
experimental studies of the $K_{L} \rightarrow\pi^{0}\gamma\gamma$ diphoton
spectrum \cite{KLpgg}. As a result of these new experimental
results, and the related theoretical analyses in Ref.~\cite{BDI,ISU,EdR},
we finally have a good control on all the components 
of $K_L \to \pi^0 \ell^+ \ell^-$ amplitudes (see Table~\ref{tab:rareK}).

In both cases the short-distance component represent 
a sizable fraction of the branching ratios. Moreover, 
electron and muon modes have different sensitivity 
to short-distance dynamics (with different relative  
weights of vector- and axial-current contributions
already within the SM).
This provides a  very powerful tool to
distinguish among different new-physics
scenarios \cite{ISU,MST}.

\subsection{Rare $K$ decays beyond the SM}

As already mentioned, the short distance nature of the 
$s \to d \nu \bar{\nu}$ transition implies a strong 
sensitivity of $K\to \pi\nu\bar\nu$ decays to possible SM 
extensions \cite{GN}. Observable deviations from the SM predictions 
are expected in many specific frameworks.
In particular, large effects are expected 
in models with non-MFV structures, such as scenarios 
with enhanced $Z$-penguins \cite{Zpenguins}, the MSSM 
with non-MFV soft-breaking terms \cite{GMSSM,IMPST,IP1}
or with $R$-parity violation \cite{deAndrea}. The effects
are much smaller in models which respect the MFV
criterion, such as the low-energy supersymmetric scenarios 
analysed in Ref.~\cite{IMPST,MSSM}, or 
the little-Higgs and large-extra-dimension models
discussed in Ref.~\cite{Little} and Ref.~\cite{Buras_extra}. 
 Present experimental data 
do not allow yet to fully explore the high-discovery potential 
of these modes. Nonetheless, it is worth to stress that 
the evidence of the $K^+\to\pi^+ \nu\bar\nu$ transition obtained at BNL \cite{BNL}
already provides highly non-trivial constraints on the realistic scenarios 
with large new sources of flavour mixing.
 
As illustrated in  Fig.~\ref{fig:stopvschar}, within the pessimistic
framework of MFV the possible deviations from the SM in 
$K \to \pi \nu\bar\nu$ are highly correlated to the spectrum 
of the new degrees of freedom. For this reason, even within MFV models 
precise measurements of these modes would be very valuable 
(allowing for very stringent tests of the theory). 
In presence of non-MFV structures, the two $K \to \pi \nu\bar\nu$ modes 
are usually the most sensitive probes of new sources of flavor symmetry 
breaking which also violates the $SU(2)_L$ gauge symmetry (such as the up-type trilinear 
terms in the MSSM, see Fig.~\ref{fig:Kpnn}). Within these general 
frameworks, significant new information can also be extracted from the 
$K_L \to \pi^0 \ell^+ \ell^-$ modes \cite{Zpenguins,IMPST}.

In summary, it is fair to say that the four rare $K$ decay modes
in Table~\ref{tab:rareK}, and particularly the two neutrino modes, 
are a mandatory ingredient for a deeper and model-independent study of the flavor
problem in the quark sector.

\section{Minimal Lepton Flavor Violation}

Since the observed neutrino mass parameters are not 
described by the SM Yukawa interaction in Eq.~(\ref{eq:LY}),
the formulation of a MFV hypothesis for the lepton 
sector is not straightforward.
A proposal based on the assumption that the breaking 
of total lepton number (LN) and lepton flavor are decoupled in the underlying theory
has recently been presented in Ref.~\cite{MLFV},
and further analysed in Ref.~\cite{CG}.
  
Two independent  MLFV scenarios have been identified. 
They are characterized by the different status assigned to
the effective Majorana mass matrix $g_\nu$  appearing as coefficient of 
the $|\Delta L| = 2$ dimension-five operator in the low 
energy effective theory~\cite{Weinberg:1979sa}:
\beq
\cL_{\rm eff}^{\nu}  = -\frac{1}{ \Lambda_{\rm LN}}\,g_\nu^{ij}(\bar
L^{ci}_L\tau_2 H)(H^T\tau_2L^j_L)  {\rm ~+~h.c.}
\eeq
In the truly minimal scenario (dubbed {\em minimal field content}),  
$g_\nu$ and the charged-lepton
Yukawa coupling ($Y_E$) are assumed to be the only irreducible sources 
of breaking of $SU(3)_{L_L}\times SU(3)_{E_R}$, the lepton-flavor 
symmetry of the low-energy theory. 
This strong assumption does not hold in many realistic underlying 
theories with heavy right-handed neutrinos. For this reason, 
a second scenario (dubbed {\em extended field content}), 
with heavy right-handed neutrinos and a larger 
lepton-flavor symmetry group, $SU(3)_{L_L}\times SU(3)_{E_R} \times O(3)_{\nu_R}$,
has also been considered. 
In this extended scenario, the most natural and 
economical choice about the symmetry-breaking terms is the identification 
of the two Yukawa couplings, $Y_\nu$ and $Y_E$, 
as the only irreducible symmetry-breaking structures 
(in close analogy with the quark sector). In this context,
$g_\nu \sim  Y_\nu^T Y_\nu$ and the 
LN-breaking  mass term of the heavy 
right-handed neutrinos is flavor-blind 
(up to Yukawa-induced corrections):
\beqa
\cL_{\rm heavy}  &=&  -\frac{1}{2} M_\nu^{ij}\bar \nu^{ci}_R\nu_R^j {\rm ~+~h.c.} 
\qquad M_\nu^{ij}=M_\nu  \delta^{ij}  \no\\
\cL^{\rm ext}_{Y}  &=&  \cL_{Y}  +i Y_\nu^{ij}\bar\nu_R^i(H^T \tau_2L^j_L) {\rm ~+~h.c.} 
\eeqa

The basic assumptions of these scenarios are
more arbitrary are less phenomenologically-driven 
than in the quark sector. Nonetheless, the formulation of 
an EFT based on these assumptions is still very useful.
As I will briefly illustrate in the next section, it allows us to 
address in a very general way the following fundamental 
question: how can we detect the presence of new irreducible 
(fundamental) sources of LF symmetry breaking? 

\section{LF violation: $\mu$ vs $\tau$ decays}

Using the MLFV-EFT approach, one can easily 
demonstrate that --in absence of new sources of LF violation--
visible FCNC decays of $\mu$ and $\tau$ can occur only 
if there is a large hierarchy between $\LLFV$
(the scale of  new degrees of freedoms carrying LF) 
and $\LLN \sim M_\nu$ (the scale of total LN violation) \cite{MLFV}.
This condition is indeed realized within the explicit 
extensions of the SM widely discussed in the literature 
which predict sizable LF violating effects in charged 
leptons (see e.g.~Ref.~\cite{Barbieri}).

\begin{figure}[t]
\begin{center}
\includegraphics[width=80mm]{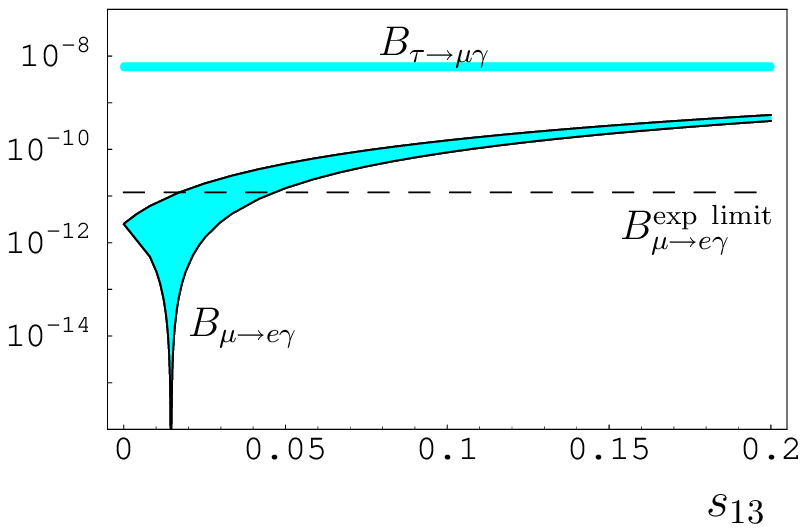}
\caption{\label{fig:MLFV}  
$B_{\tau \to  \mu \gamma} = \Gamma(\tau \to \mu \gamma)/\Gamma(\tau \to \mu \nu \bar{\nu}) $
compared to the $\mu  \to  e \gamma$ constraint within MLFV (minimal field
content),  as a function of the neutrino mixing angle $s_{13}$ \cite{MLFV}.
The shading corresponds to different values of the phase $\delta$ 
and the normal/inverted spectrum. The NP scales have been set to
 $\Lambda_{\rm LN}/\Lambda = 10^{10}$; their variation affects 
only the overall vertical scale.}
\end{center}
\end{figure}

\begin{figure}[t]
\begin{center}
\includegraphics[width=80mm]{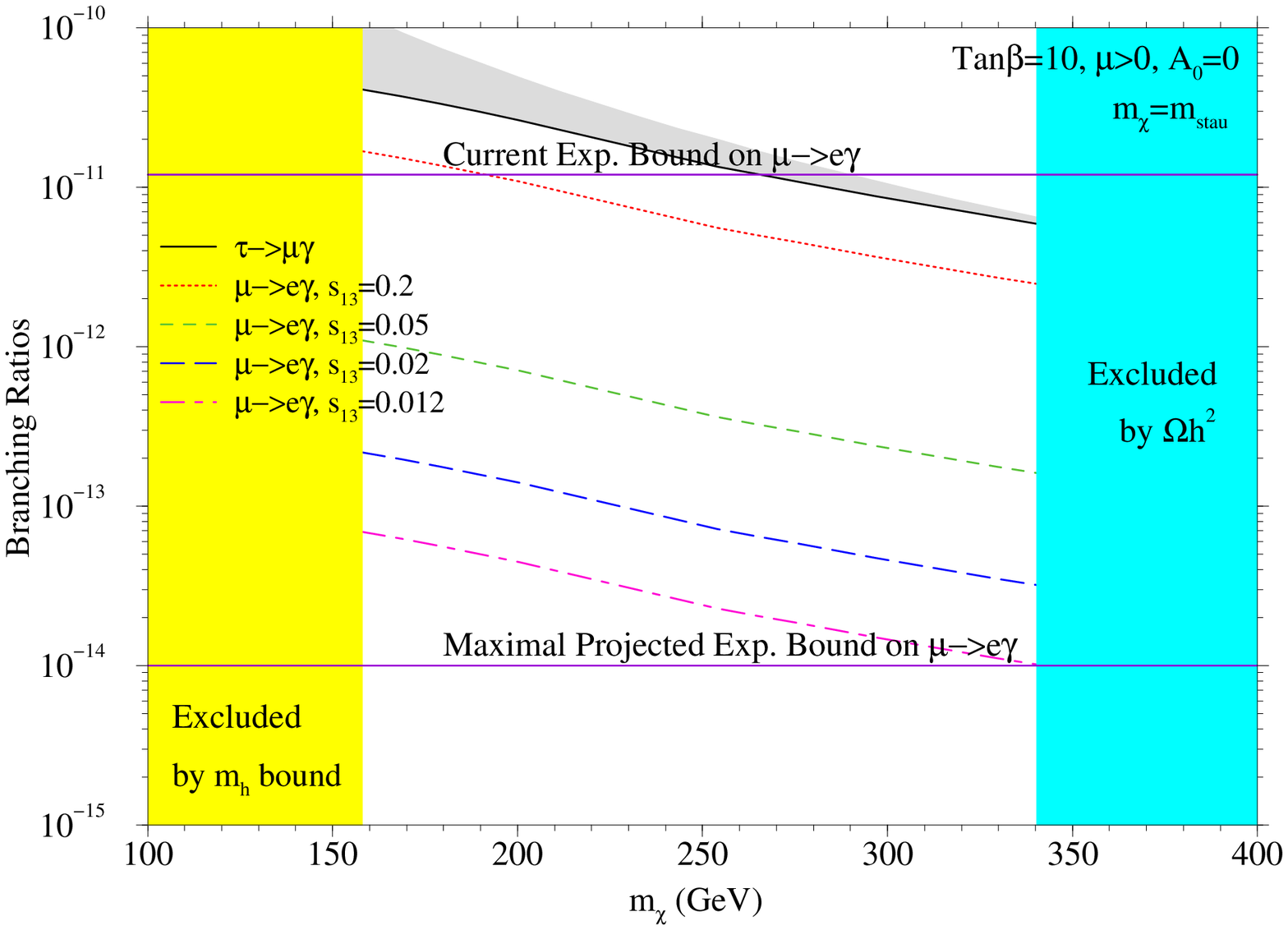}
\caption{\label{fig:profumo}  
Isolevel curves for $\cB(\mu\rightarrow e\gamma)$ and 
$\cB(\tau\rightarrow \mu\gamma)$ in the MSSM (for $\tan\beta=10$, $\mu>0$ and $A_0=0$) 
compared with the present and future experimental resolution on
$\cB(\mu\rightarrow e\gamma)$ \cite{profumo}.}
\end{center}
\end{figure}

More interestingly, the EFT  allows us to draw unambiguous 
predictions about the relative size of LF violating decays 
of charged leptons (in terms of neutrino masses and mixing angles). 
At present, the uncertainty in the predictions for such ratios is limited 
from the poorly constrained value of the $1$--$3$ mixing angle in the neutrino
mass matrix ($s_{13}$) and, to a lesser extent, 
from the neutrino spectrum ordering and the CP violating phase $\delta$. 
One of the  clearest consequences from the phenomenological
point of view is the observation that  if $s_{13} \gsim 0.1$ there 
is no hope to observe $\tau \to \mu \gamma$ at future accelerators 
(see Fig.~\ref{fig:MLFV}). This happens because the stringent
constraints from $\mu \to e \gamma$ already forbid too low values 
for the effective scale of LF violation. In other words,
in absence of new sources of LF violation the most sensitive 
FCNC probe in the lepton sector is $\mu \to e \gamma$.
This process should indeed be observed at  MEG \cite{MEG}
for very realistic values of the new-physics scales $\LLFV$ 
and $\LLN$.  

The expectation of a higher NP sensitivity of 
$\mu \to \mu \gamma$ with respect to 
$\tau \to \mu \gamma$
(taking into account the corresponding experimental resolutions) 
is confirmed in several realistic NP frameworks.
This happens for instance in the MSSM (see Fig.~\ref{fig:profumo})
with the exception of specific corners 
of the parameter space~\cite{hisano}.

\section{LF universality tests in $K_{\ell 2}$ and $\pi_{\ell 2}$}

An alternative phenomenological tool to search for new 
sources of LF violation is provided by precise 
LF universality tests in charged-current 
meson decays. In particular, the ratios 
\beq
R_P^{\mu/e} = \frac{ \cB(P\to \mu \nu) }{ \cB(P \to e \nu)}
\eeq
can be predicted with excellent accuracies in the SM, 
both for $P=\pi$ (0.02\% accuracy \cite{Marciano}) and $P=K$ 
(0.04\% accuracy \cite{Marciano}), allowing for some 
of the most significant tests of LF universality. 

Within MLFV models, it is easy to realise that these 
ratios cannot be modified at appreciable levels:
the presence of the charged-lepton Yukawa coupling ($Y_E$) in the 
operators involving the right-handed lepton fields
makes their contribution safely negligible. However, 
as recently pointed out in Ref.~\cite{kl2}, this 
suppression can be avoid in realistic non-MFV 
scenarios which can occur in specific 
supersymmetric frameworks.

The key ingredients which allows visible non-SM 
contributions in $R_P^{\mu/e}$ within the MSSM are:
\begin{enumerate}
\item[i)] large values of $\tan\beta$ (the ratio of the two Higgs 
vacuum expectation values), such that the overall normalization of 
$Y_E$ --and correspondingly the $H^{\pm}$-exchange 
contribution to $P\to \ell \nu$-- is enhanced;
\item[ii)] large mixing angles in the right-slepton sector,
such that the $P\to \ell_i \nu_{j}$ rate (with $i\not=j$)
becomes non negligible. 
\end{enumerate}
In the most favorable scenarios,
the deviations from the SM could reach $\sim 1\%$ in the 
$R_K^{\mu/e}$ case \cite{kl2} (not far from the present 
experimental resolution \cite{kl2_exp}) 
and $\sim {\rm few} \times 10^{-4}$ in the $R_\pi^{\mu/e}$
case. In the pion case the effect is quite below the 
present experimental resolution~\cite{pl2_exp}, 
but could well be within the reach of the new
generation of high-precision $\pi_{\ell 2}$ 
experiments planned at TRIUMPH and at PSI.
  
In principle, larger
violations of LF universality are expected in 
$B \to \ell \nu $ decays, with $\cO(10\%)$ deviations 
from the SM in $R_B^{\mu/\tau}$ and even order-of-magnitude 
enhancements in $R_B^{e/\tau}$~\cite{IP2}.
However, the difficulty of precision measurements 
of the highly suppressed $B \to e/\mu~ \nu$ modes
makes these non-standard effects undetectable (at least at present). 

Similarly to the FCNC decays
discussed in the previous sections, 
also for the LF universality tests 
the low-energy systems ($K_{\ell 2}$ and $\pi_{\ell 2}$)
offer a unique opportunity 
in shedding light on physics beyond the Standard Model: 
the smallness of NP effects is more than compensated (in terms of NP sensitivity) 
by the excellent experimental resolution and the 
good theoretical control.

\section*{Acknowledgments}
It is a pleasure to thank the organizers 
of FPCP2006 for the invitation to this 
enjoyable and interesting meeting.

\bigskip 

\end{document}